\documentclass[a4paper,11pt]{article}
\pdfoutput=1 % if your are submitting a pdflatex (i.e. if you have
\usepackage{jheppub} % for details on the use of the package, please
\usepackage[T1]{fontenc} % if needed
\usepackage{amssymb,amsmath,graphicx,amscd,xcolor,amsthm}
\usepackage{braket}
\usepackage{verbatim} %comment

\def\k{\mathbf{k}}

\title{\boldmath Vacuum fluctuations of a scalar field near a reflecting boundary and their effects on the motion of a test particle}
\author[a]{G. H. S. Camargo,}
\author[a]{V. A. De Lorenci,}
\author[b]{C. C. H. Ribeiro,}
\author[a]{F. F. Rodrigues,}
\author[c]{ and M. M. Silva}

\affiliation[a]{Instituto de F\'{\i}sica e Qu\'{\i}mica,  Universidade Federal de Itajub\'a,\\
Itajub\'a, Minas Gerais 37500-903, Brazil}
\affiliation[b]{Instituto de F\'{\i}sica de S\~ao Carlos, Universidade de S\~ao Paulo,\\ S\~ao Carlos, S\~ao Paulo 15980-900, Brazil}
\affiliation[c]{Dipartimento di Fisica, Universit\`a `La Sapienza',\\  I-00185 Roma, Italy}

\emailAdd{guilhermehenrique@unifei.edu.br}
\emailAdd{delorenci@unifei.edu.br}
\emailAdd{caiocesarribeiro@ifsc.usp.br}
\emailAdd{fernanda-fr@unifei.edu.br}
\emailAdd{silva.malumaira@gmail.com}

\date\today

\abstract{
The contribution from quantum vacuum fluctuations of a real massless scalar field to the motion of a test particle that interacts with the field in the presence of a perfectly reflecting flat boundary is here investigated. 
There is no quantum induced dispersions on the motion of the particle when it is alone in the empty space.  However, when a reflecting wall is introduced, dispersions occur with magnitude dependent on how fast the system evolves between the two scenarios.
A possible way of implementing this process would be by means of an idealized sudden switching, for which the transition occurs  instantaneously. 
Although the sudden process is a simple and mathematically convenient idealization it brings some divergences to the results, particularly at a time corresponding to a round trip of a light signal between the particle and the wall. 
It is shown that the use of smooth switching functions, besides regularizing such divergences, enables us to better understand the behavior of the quantum dispersions induced on the motion of the particle. 
Furthermore, the action of modifying the vacuum state of the system leads to a change in the particle energy that depends on how fast the transition between these states is implemented.  
Possible implications of these results to the similar case of an electric charge near a perfectly conducting wall are discussed. 
}


\begin{document} 
\maketitle
\flushbottom

\section{Introduction}
Quantum vacuum fluctuations of fields in the presence of boundaries or in nontrivial spacetimes are known to be responsible for several predicted effects in different branches of physics. The most famous one is the well known and experimentally tested Casimir effect \cite{lamoreaux2005,klimchitskaya2009} (see Ref. \cite{liu2016} for a current application). Just to mention few other examples, local thermal behavior of a massive scalar field near a reflecting wall leads to the existence of bounds on the coupling curvature parameter \cite{delorenci2015xi,moreira2017}, and in the context of semiclassical gravity we report a sort of quantum dragging effect \cite{delorenci2004} in a spacetime of a cosmic string presenting a dislocation along its axial axis.

Another interesting effect powered by quantum vacuum fluctuations is the occurrence of dispersions of the velocity of a test particle interacting with a quantum field in the neighborhood of a boundary \cite{ford2004}.
In the case of electric charge near a perfectly reflecting flat wall, it was shown \cite{ford2004,ford2005} that quantum dispersions are induced on the motion of a charged test particle due to the modification of the vacuum state of the system, as compared to the empty space vacuum state (the Minkowski vacuum state). When an idealized model with a sudden transition between the scenarios of a particle in the empty space and a particle near a perfectly reflecting wall is assumed, it results that the dispersions present a typical divergence at the wall position which is linked to the boundary condition. Additionally, another divergence occurs after an interval of time corresponding to a round trip of a light signal between the particle and the wall. An interesting behavior of this system is the reported residual effect at late times, which may be connected to an energy conservation law \cite{ford2004}. Several aspects about this system have been discussed  so far \cite{hongwei2004,hongwei2006,seriu2008,seriu2009,parkinson2011,delorenci2016}, including the study of Brownian motion in curved spacetime \cite{bessa2009}. Particularly, it was shown \cite{delorenci2016} that smoothing the transition between the above mentioned states of the system leads to a natural regularization of all divergences previously reported. Moreover, the late time behavior of the dispersions appears to be dependent on the properties of the implemented switching. For instance, in the model examined in Ref. \cite{delorenci2016} no residual effect survives.

Simplified (1+1)-dimensional systems based on a test particle interacting with a massless scalar field in the presence of a point-like mirror \cite{delorenci2014} and inside a cavity with one mobile wall \cite{bartolo2015} were also examined. In the case of a point-like mirror it was shown that when a sudden switching is implemented the dispersion of the particle velocity is affected by the same kind of divergences as those appearing in the case of an electric charge interacting with the electromagnetic field. In such simplified model implementing a mechanism of distance fluctuations is enough to regularize the divergences. However, late time behavior could not be discussed in such investigation as it was reported to be beyond the limit of applicability of the model. 
Still in 1+1 dimensions, the motion of a mirror with an internal harmonic oscillator coupled to a scalar field was investigated in Ref. \cite{wang2014}.

Quantum dispersions induced by vacuum fluctuations of a scalar field in an analog model for Friedmann-Robertson-Walker (FRW) spatially flat spacetime were recently studied in the context of an expanding Bose-Einstein condensate \cite{bessa2017}. As a result it was claimed that implementing a time dependent scale factor is enough to regularize the divergences that appear when one reflecting boundary is present, and thus it would play the same role of a smooth switching function in the regularization process. However, only late time regime was discussed, for which no singularities linked to time intervals are known to occur. 

In this paper the motion of a test particle interacting with a scalar field in the presence of a perfectly reflecting wall in 3+1 dimensions is studied. The results show that the behavior of the dispersions are quite similar to the case of an electric charge interacting with the electromagnetic field. The residual effect reported in the electromagnetic case also appears in the scalar field model when the sudden process is considered. However, such effect is highly dependent on the adopted switching. Two convenient smooth switching scenarios are implemented and their implications in the behavior of the dispersion of the particle velocity are investigated. 
Furthermore, it is discussed how the rate of transition between the vacuum states affects the amount of energy exchanged with the particle. Particularly, late time behavior analysis shows that subvacuum effects are responsible for a suppression of a certain amount of kinetic energy of the particle, whose magnitude depends on the rate of transition.
Implications of these results to the recently investigated case of an electric charge near a conducting wall are also addressed.  

In the next section basic aspects about the interaction between a test particle and a real massless scalar field are discussed. The quantization of the system in the presence of a reflecting boundary is shortly presented in Sec. \ref{quantum}, which includes the formal expression for the calculation of the dispersion of the particle velocity. Thus, the following section deals with the derivation of the dispersions when the different scenarios are assumed. The first one consists of implementing a sudden switching, while in the others it is assumed a smooth transition between the two distinct states of the system.  
A discussion about the kinetic energy of the particle is done in Sec. \ref{e-kinetic}.
Final remarks are presented in Sec. \ref{final}, including a short discussion about the implications of our results to the similar system based on electromagnetic interaction. An appendix presenting the detailed derivation of the dispersions in the scenario of one of the smooth switching functions closes this work. Units are such that $\hbar=c=1$.

\section{Interaction between a test particle and a scalar field}
In the nonrelativistic regime, a point-like particle of mass $m$ and scalar charge $g$, interacting with a massless scalar field $\phi\left({\bf x},t\right)$ in 3+1 dimensions, has its motion governed by the Lorentz-like force \cite{delorenci2014} $m (d v_i/dt)=-g\partial_i\phi\left({\bf x},t\right)$, where the subindex $i$ runs from 1 to 3, corresponding to the three Cartesian coordinates $x,y,z$. 
If we restrict our study to the case where the particle position does not appreciably vary in time \cite{ford2004,ford2005}, the particle velocity reads
\begin{equation}
v_i=-\frac{g}{m}\frac{\partial}{\partial x_i}\int_0^\tau\phi\left({\bf x},t\right)d t.
\label{int6}
\end{equation}
This expression assumes a sudden switching regime, as the interaction is sharply limited to the interval  $0 < t < \tau$. In other words, the interaction is turned on at $t=0{\rm s}$ and turned off after a time $\tau$. Henceforth we will refer to $\tau$ as the measuring time -- the interval of time during which the particle is effectively interacting with the field.  This description is indeed quite convenient as an idealized model. Nevertheless, transient effects can lead to important contributions, motivating the search for a more complete model. 

A more realistic scheme can be implemented by introducing a switching function $F_{\tau}(t)$ that smoothly connects the two distinct regimes of the system. For instance, when only classical aspects are considered, the switching could provide a continuous way connecting the cases when the particle is free of interaction and when it effectively interacts with the field. Mathematically, Eq. (\ref{int6}) should be amended as
\begin{equation}
v_i=-\frac{g}{m}\frac{\partial}{\partial x_i}\int_{-\infty}^{\infty}\phi\left({\bf x},t\right) F_{\tau}(t) dt,
\label{int6F}
\end{equation}
where the following normalization must be observed,
\begin{equation}
\frac{1}{\tau}\int_{-\infty}^\infty F_\tau\left(t\right)dt=1.
\label{int25}
\end{equation}

The sudden process assumed in Eq. (\ref{int6}) is now understood as a limiting case for which $F_{\tau}(t) \rightarrow \Theta(t)\Theta(\tau-t)$, where $\Theta(t)$ represents the unit step function, which is equal to 0 for $t<0$, and 1 for $t\ge 1$.

\section{Quantizing the system}
\label{quantum}
The system we are going to study consists of a test particle near a reflecting flat wall which is placed at $z=0$ during a certain interval of time $\tau$. We are interested in studying the dispersion of the particle velocity when the quantum scalar field is prepared in its vacuum state, which differs from the Minkowski vacuum state due to the presence of the wall.
In the realm of quantum field theory, the scalar field is described by an operator $\phi\left({\bf x},t\right)$ that can be obtained by means of the standard scheme of quantization \cite{birrel1982}. 
The expansion of the field in terms of normal modes gives $\phi\left({\bf x},t\right)=\int d^3k \left[ a_{\bf k} u_{\bf k}\left({\bf x},t\right)+ a^\dagger_{\bf k} u_{\bf k}^*\left({\bf x},t\right)\right],$
where $ a_\k$ and $ a^\dagger_\k$ are, respectively, the annihilation and the creation operators that satisfy the commutation relation $[ a_{\bf k}, a^\dagger_{\bf k'}]=\delta (\k - \k')$. 
The normal modes  are obtained by solving the wave equation $\Box u_{\bf k}\left({\bf x},t\right) =0$, where $\Box$ is the d'Alembertian operator in Minkowski spacetime.  
The presence of the reflecting wall is translated in terms of the boundary conditions over the modes as $u_{\bf k}\left(x,y,z=0,t\right)=0$.   
Following the above prescription we obtain the Klein-Gordon normalized modes, 
\begin{equation}
u_{\bf k}\left({\bf x},t\right)=\frac{1}{\sqrt{4\pi^3\omega}}\,{\rm e}^{-i(\omega t - k_x x - k_y y)}\sin{\left(k_z z\right)},
\label{int13}
\end{equation}
where $\omega=\left|{\bf k}\right|$.

The vacuum state is defined by $ a_\k|0\rangle =0$, which implies that $\braket{ a_\k}\doteq \braket{0| a_\k|0}=0$ and $\braket{ a^\dagger_\k}=0$. Thus, when the scalar field is a pure quantum quantity, its vacuum expectation value is also zero, $\braket{\phi\left({\bf x},t\right)}=0$. Hence, the dispersion of the quantum scalar field is simply  $\braket{(\Delta\phi\left({\bf x},t\right))^2} = \braket{\phi\left({\bf x},t\right)^2} -\braket{\phi\left({\bf x},t\right)}^2=\braket{\phi\left({\bf x},t\right)^2}$. It is worth mentioning that in the absence of thermal effects, even in the case of having a classical contribution, only the quantum part of the total field contributes to the dispersion. This is because the classical contribution is a c-number that would identically cancel in the above subtraction.  

The particle velocity is now taken as a Hermitian operator whose vacuum expectation value is zero, $\braket{ v_i}=0$. Its dispersion is  different from zero, and is given by $\braket{(\Delta v_i)^2} = \braket{( v_i -\braket{ v_i})^2} = \braket{ v_i{}^2} - \braket{ v_i}^2 = \braket{ v_i{}^2}$. After squaring Eq. (\ref{int6F}), symmetrizing the resulting expression with respect to the product of the fields, and taking its vacuum expectation value, we obtain
\begin{equation}
\braket{(\Delta v_i)^2}=\frac{g^2}{2m^2}\left[\frac{\partial}{\partial x_i}\frac{\partial}{\partial x_i^\prime}\int_{-\infty}^\infty\int_{-\infty}^\infty G^{{}^{(1)}}(\textbf{x},t;\textbf{x}^\prime,t^\prime)F_\tau\left(t\right)F_\tau\left(t'\right)d td t^\prime\right]_{\textbf{x}^\prime=\textbf{x}},
\label{int10a}
\end{equation}
where the Hadamard two-point function $G^{{}^{(1)}}(\textbf{x},t;\textbf{x}^\prime,t^\prime) = \braket{\phi\left(\textbf{x},t\right)\phi\left(\textbf{x}^\prime,t^\prime\right)}+\braket{\phi\left(\textbf{x}^\prime,t^\prime\right)\phi\left(\textbf{x},t\right)}$ was identified. Using the above results it is straightforward to show that  $G^{{}^{(1)}}(\textbf{x},t;\textbf{x}',t') = G^{{}^{(1)}}_{{}_0}(\textbf{x},t;\textbf{x}',t') + G^{{}^{(1)}}_{{}_R}(\textbf{x},t;\textbf{x}',t')$, where $G^{{}^{(1)}}_{{}_0}(\textbf{x},t;\textbf{x}',t')$ is the empty space Hadarmard two-point function (Minkowski propagator), while
\begin{equation}
G^{{}^{(1)}}_{{}_R}(\textbf{x},t;\textbf{x}',t') = \frac{1}{2\pi^2\left[\left(t-t^\prime\right)^2 - (x-x')^2-(y-y')^2-(z+z')^2\right]}
\label{3.19b}
\end{equation}
is the renormalized propagator  \cite{birrel1982}.  

In what follows we will use the renormalized version of Eq. (\ref{int10a}). It is worth mentioning that empty space zero-point fluctuations, linked to $G^{{}^{(1)}}_{{}_0}(\textbf{x},t;\textbf{x}',t')$,  do not cause observable effects over the motion of a particle \cite{johnson2002}.  

Renormalized vacuum fluctuations of a massless scalar field near a reflecting boundary are well known in the literature. Particularly, it was shown that the divergent behavior of the energy density ($T_{tt}$) at $z=0$ can be regularized by introducing a nonzero position uncertainty to the boundary \cite{ford1998} (see also Ref. \cite{armata2015} for the case of a cavity with a mobile boundary).

\section{Dispersions of the particle velocity}
\label{sec-dispersions}
In what follows we will study the dispersions of the particle velocity in two different scenarios.  The first one is the sudden switching, for which a reflecting wall is assumed to be instantaneously placed near the particle, and after a time $\tau$ it is instantaneously removed. Next we will discuss two possible ways of implementing a smooth switching, bringing more reality to the description of the system.

\subsection{Sudden switching scenario}
 In this first scenario, one can clearly distinguish between two different situations. If $t < 0$ or $t > \tau$ the particle is in the empty space, while in the interval $0 < t < \tau$ the particle is in the neighborhood of a flat reflecting wall.  The dispersions can be obtained by assuming the idealized case for which $F_{\tau}(t)$ is given by $\Theta(t)\Theta(\tau-t)$. Thus, if we introduce Eq.~\eqref{3.19b} in Eq.~\eqref{int10a}, use the identity $\int_{0}^{\tau}dt\int_{0}^{\tau}dt'f(|t-t'|)=2\int_{0}^{\tau}(\tau-\eta)f(\eta)d\eta$, and operate with the derivatives, we obtain
\begin{equation}
\braket{\left(\Delta v_\perp\right)^2}_{{}_{\mathrm{Sudden}}}=-\frac{g^2}{16\pi^2m^2z^2}\left[\frac{2\tau^2}{\left(4z^2-\tau^2\right)}+\frac{\tau}{z}\ln{\left|\frac{2z+\tau}{2z-\tau}\right|}\right]
\label{3.7}
\end{equation}
for the dispersion of the velocity in the direction perpendicular ($z$ direction) to the wall, and 
\begin{equation}
\braket{\left(\Delta v_\parallel\right)^2}_{{}_{\mathrm{Sudden}}}=-\frac{g^2\tau}{32\pi^2m^2z^3}\ln{\left|\frac{2z+\tau}{2z-\tau}\right|}
\label{3.9}
\end{equation}
for the dispersion of the velocity in the direction parallel ($x$ and $y$ directions) to the wall. When some parameters are conveniently chosen, the above results coincide with the dispersions calculated for a free particle near a reflecting boundary in an analog FRW spacetime \cite{bessa2017}.

Notice that in both directions the dispersions diverge at $z=0$ and $\tau = 2z$. These divergences were already reported in the literature for similar systems \cite{ford2004,delorenci2014,delorenci2016} and are related to idealized boundary and switching conditions. We will see that implementing a smooth switching is enough to regularize theses infinities. Another important aspect exhibited by the dispersions is a residual effect at late times, i.e.,
\begin{equation}
\lim_{\tau\to\infty}\braket{\left(\Delta v_\perp\right)^2}_{{}_{\mathrm{Sudden}}}=\lim_{\tau\to\infty}\braket{\left(\Delta v_\parallel\right)^2}_{{}_{\mathrm{Sudden}}}=-\frac{g^2}{8\pi^2 m^2 z^2}.
\label{3.10}
\end{equation}

A residual dispersion also appears in the case of an electron near a flat and perfectly conducting wall when sudden approximation is used \cite{ford2004}, and such result was suggested to be linked with the energy conservation law. The late time behavior of the dispersions is closely related to the switching interval of time (hereafter denoted by $\tau_s$) between the two different regimes of the system. In a sudden switching scenario $\tau_s$ is just zero. 

\subsection{Smooth switching scenario I}
\label{sss}
\begin{figure}
\centering
\includegraphics[scale=0.33]{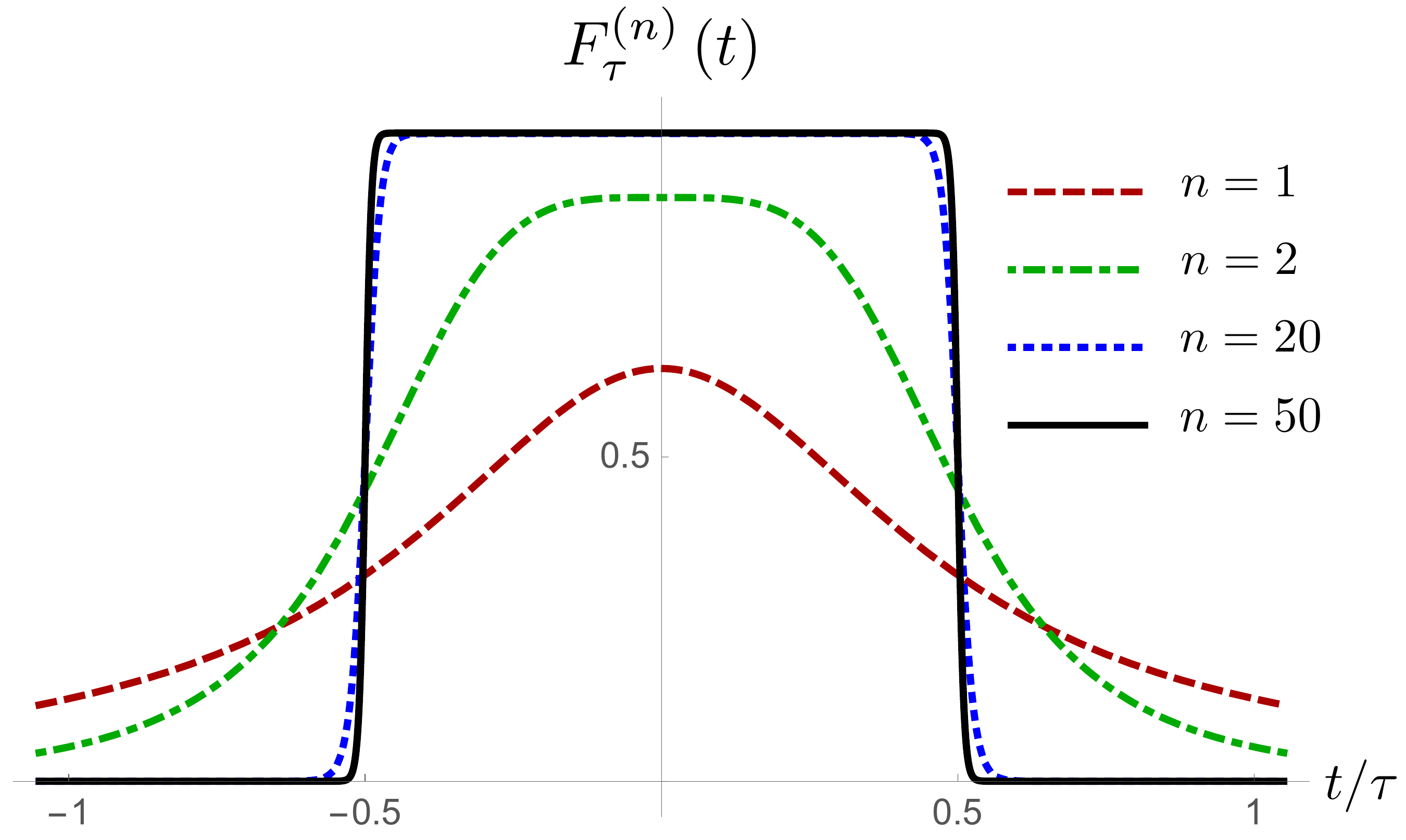}
\caption{Behavior of $F_{\tau}^{(n)}\left(t\right)$ given by Eq.~\eqref{sw1} for some representative values of $n$. The sudden switching regime is achieved when $n \rightarrow \infty$. }
\label{fnt}
\end{figure}

In order to bring more reality to the description of the system under study we can implement a smooth switching by selecting one of the functions in the sequence of functions given by \cite{delorenci2016} 
\begin{equation}
F_\tau^{\left(n\right)}\left(t\right) =\frac{c_n}{1+(2t/\tau)^{2n}},
\label{sw1}
\end{equation}
where $c_n=\left(2n/\pi\right)\sin{\left(\pi/2n\right)}$, with integer $n$, guarantees the normalization required by Eq. (\ref{int25}). The value of the parameter $n$ can be chosen accordingly to the physical requisites of the system. It should be noticed that the step function like behavior described by the sudden approximation is here recovered as $n \rightarrow \infty$. All these aspects are depicted in Fig. \ref{fnt}.
Formally, the expression for the switching interval of time $\tau_{s}$ can be obtained by taking the difference between the closest points of maximum curvature of $F^{(n)}_\tau(t)$, which are given by the zeros of the third time-derivative of this function. Notice that $\tau_s$ gives a measure of how fast the change between the different regimes is described by  $F_\tau^{\left(n\right)}(t)$. For large values of  $n$ it can be shown \cite{delorenci2016} that 
\begin{equation}
\frac{\tau_s}{\tau} \approx \frac{1}{2n}\ln\left(2+\sqrt{3}\right).
\label{ts_approx}
\end{equation} 
Notice that in the regime of $\tau\rightarrow\infty$ we will have an infinite switching time, and that corresponds to an adiabatic process.

Now, in order to obtain the dispersions of the particle velocity in this new scenario we introduce the smooth switching by implementing  $F_\tau(t) \rightarrow F_\tau^{\left(n\right)}(t)$ in the renormalized version of Eq.~\eqref{int10a}. Thus, operating with the derivatives, and using the result  (see the appendix in Ref. \cite{delorenci2016})
\begin{equation}
\begin{aligned}
&\int_{-\infty}^\infty\int_{-\infty}^\infty\frac{\left(t-t^\prime\right)^j+\left(2z\right)^k}{\left[\left(2z\right)^2-\left(t-t^\prime\right)^2\right]^l}\frac{dt}{1+\left(2t/\tau\right)^{2n}}\frac{dt^\prime}{1+\left(2t'/\tau\right)^{2n}}=\\
&\frac{\pi^2}{n^2}\left(\frac{\tau}{2}\right)^{2+j-2l}\sum_{p=0}^{n-1}\sum_{q=n}^{2n-1}\psi_{n,p}\psi_{n,q}\frac{\left(2z\right)^k\left(2/\tau\right)^j+\left(\psi_{n,q}-\psi_{n,p}\right)^j}{\left[\left(2z\right)^2\left(2/\tau\right)^2-\left(\psi_{n,p}-\psi_{n,q}\right)^2\right]^l},
\end{aligned}
\label{3.16}
\end{equation}
where $j$, $k$, $l$ are positive integers or null and $\psi_{n,p}=e^{i\frac{\pi}{2n}\left(1+2p\right)}$, the dispersion in the perpendicular direction yields
\begin{equation}
\braket{\left(\Delta v_\perp\right)^2}=-\frac{2g^2c_n^2}{m^2 n^2\tau^2}\sum_{p=0}^{n-1}\sum_{q=n}^{2n-1}\psi_{n,p}\psi_{n,q}\frac{3\left(4z/\tau\right)^2+\left(\psi_{n,p}-\psi_{n,q}\right)^2}{\left[\left(4z/\tau\right)^2-\left(\psi_{n,p}-\psi_{n,q}\right)^2\right]^3}\,,
\label{3.17}
\end{equation}
while in the parallel direction,
\begin{equation}
\braket{\left(\Delta v_\parallel\right)^2}=-\frac{2g^2c_n^2}{m^2n^2\tau^2}\sum_{p=0}^{n-1}\sum_{q=n}^{2n-1}\frac{\psi_{n,p}\psi_{n,q}}{\left[\left(4z/\tau\right)^2-\left(\psi_{n,p}-\psi_{n,q}\right)^2\right]^2}\,.
\label{3.18}
\end{equation}

\begin{figure}
\centering
\includegraphics[scale=0.4]{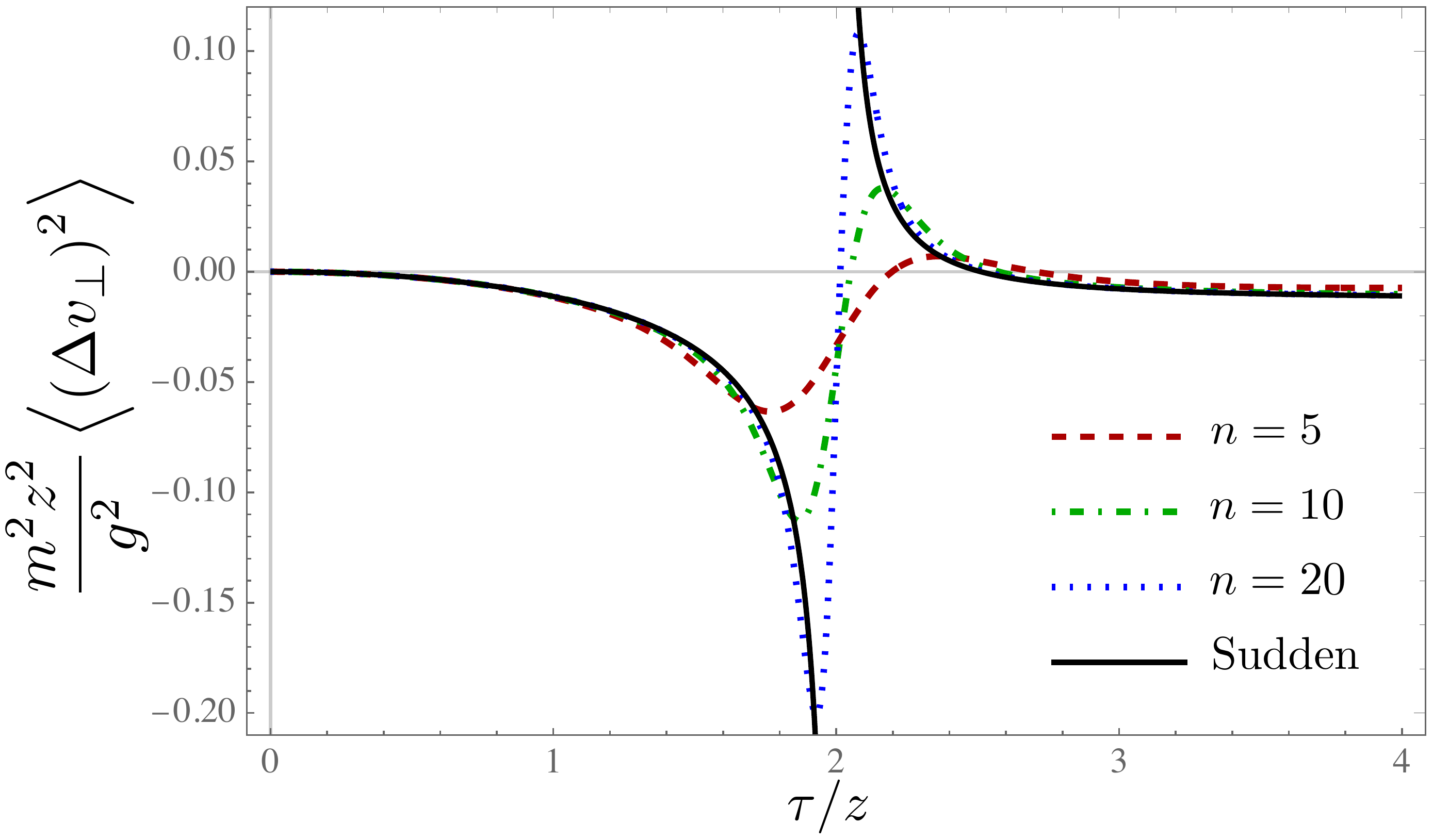}
\caption{Dispersion of the perpendicular component of the particle velocity as function of $\tau$ for the sudden switching regime (solid curve) and for some values of $n$ in the case of a smooth switching.}
\label{fig16}
\end{figure}

\begin{figure}
\centering
\includegraphics[scale=0.4]{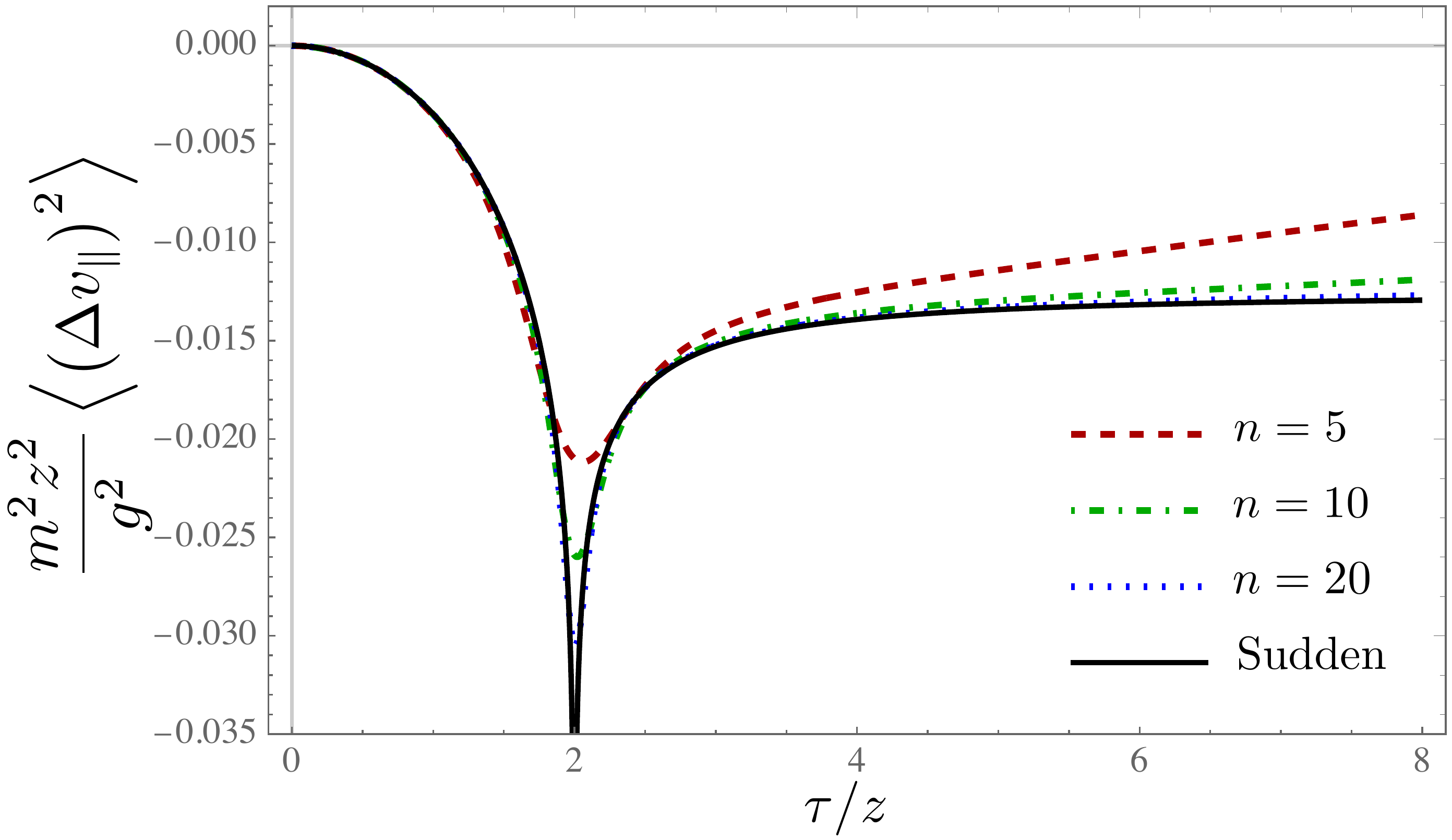}
\caption{Dispersion of the parallel component of the particle velocity as function of $\tau$ for the sudden switching regime (solid curve) and for some values of $n$ in the case of a smooth switching.}
\label{fig17}
\end{figure}

Simple inspection shows that the divergent behavior at $\tau = 2z$, that appears when the sudden switching is assumed, do not occur in this more realistic scenario. Even the divergence at $z=0$, usually linked to the Dirichlet boundary condition on the wall, is here regularized. 
Figures \ref{fig16} and \ref{fig17} depict the behavior of these dispersions for some illustrative values of $n$. Notice that the sudden switching behavior (solid curve in both figures) is more closely approached as larger values of $n$ are taken, as anticipated. %

\begin{figure}
\centering
\includegraphics[scale=0.4]{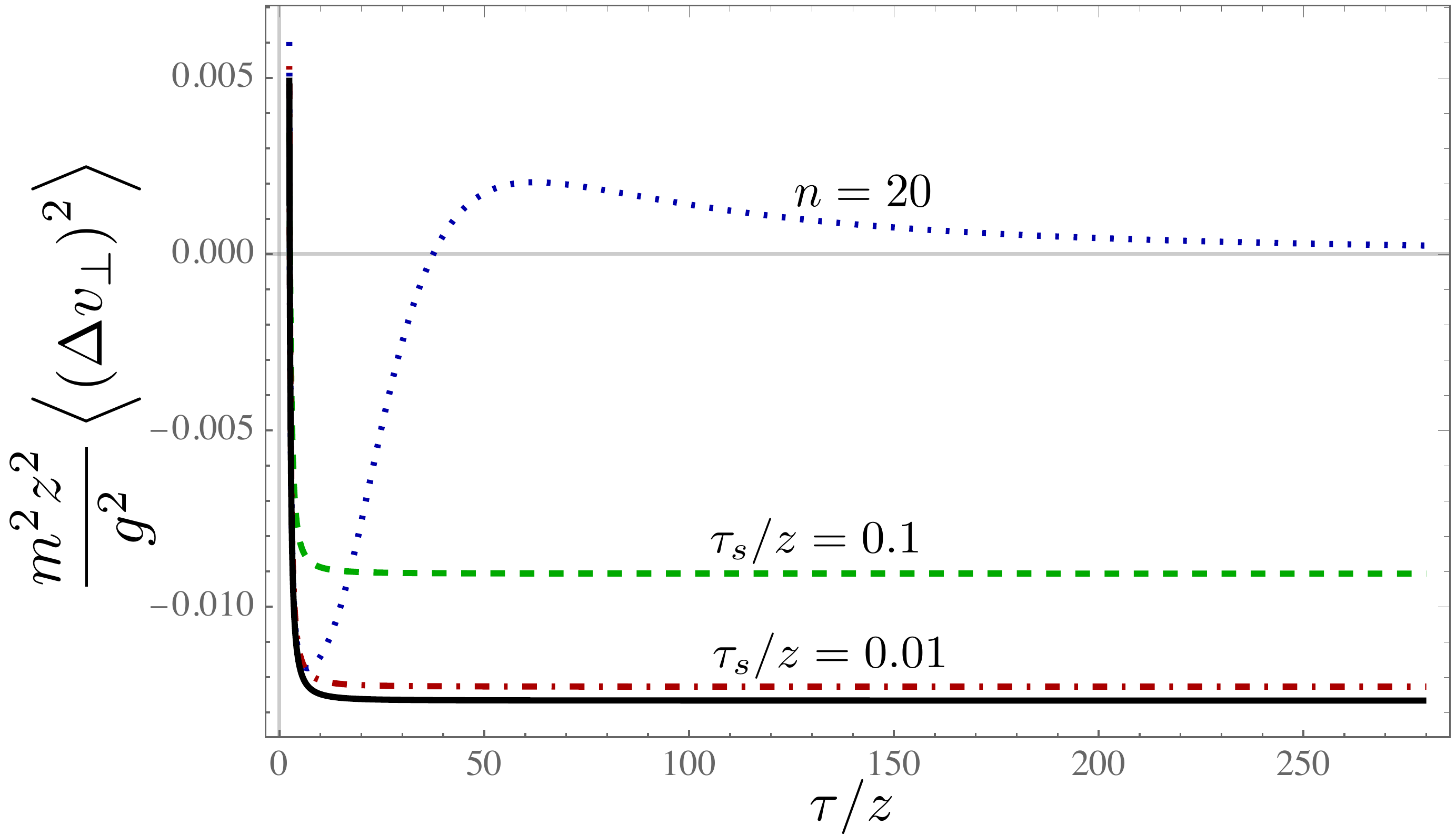}
\caption{Late time behavior of the dispersion of perpendicular component of the particle velocity when the switching $F_\tau^{(n)}(t)$ is implemented (dotted curve), and when the alternative switching $F_{\tau_s,\tau}(t)$ is implemented (dashed and dot-dashed curves). The solid curve describes the sudden switching. The curves start at $\tau/z = 2.4$ in this figure.}
\label{fig3}
\end{figure}

An important difference between smooth and sudden switchings relies on their asymptotic behavior.  As already mentioned, when the sudden switching is assumed the dispersions are governed by a function of distance $z$ in the regime of $\tau \rightarrow \infty$, and it is finite for finite z. This late time behavior is described by the solid curve in Figs. \ref{fig3} and \ref{fig5}.  However, when the smooth switching $F_\tau^{\left(n\right)}(t)$ is considered (for a finite value of $n$) the dispersions tend to zero in this same regime (see the dotted curve in the same figures). Looking more closely at the properties of this function we see that $\tau_s$ increases linearly with $\tau$, as Eq. (\ref{ts_approx}) shows. Hence, taking $\tau\rightarrow\infty$ corresponds to implementing a process for which the rate of switching is practically zero, i.e., the system performs an infinitely slowly transition between the two regimes. 
Remind that the test particle is not affected by vacuum fluctuations of a scalar field in empty space . 

\subsection{Smooth switching scenario II}
\label{sss2}
\begin{figure}
\centering
\includegraphics[scale=0.4]{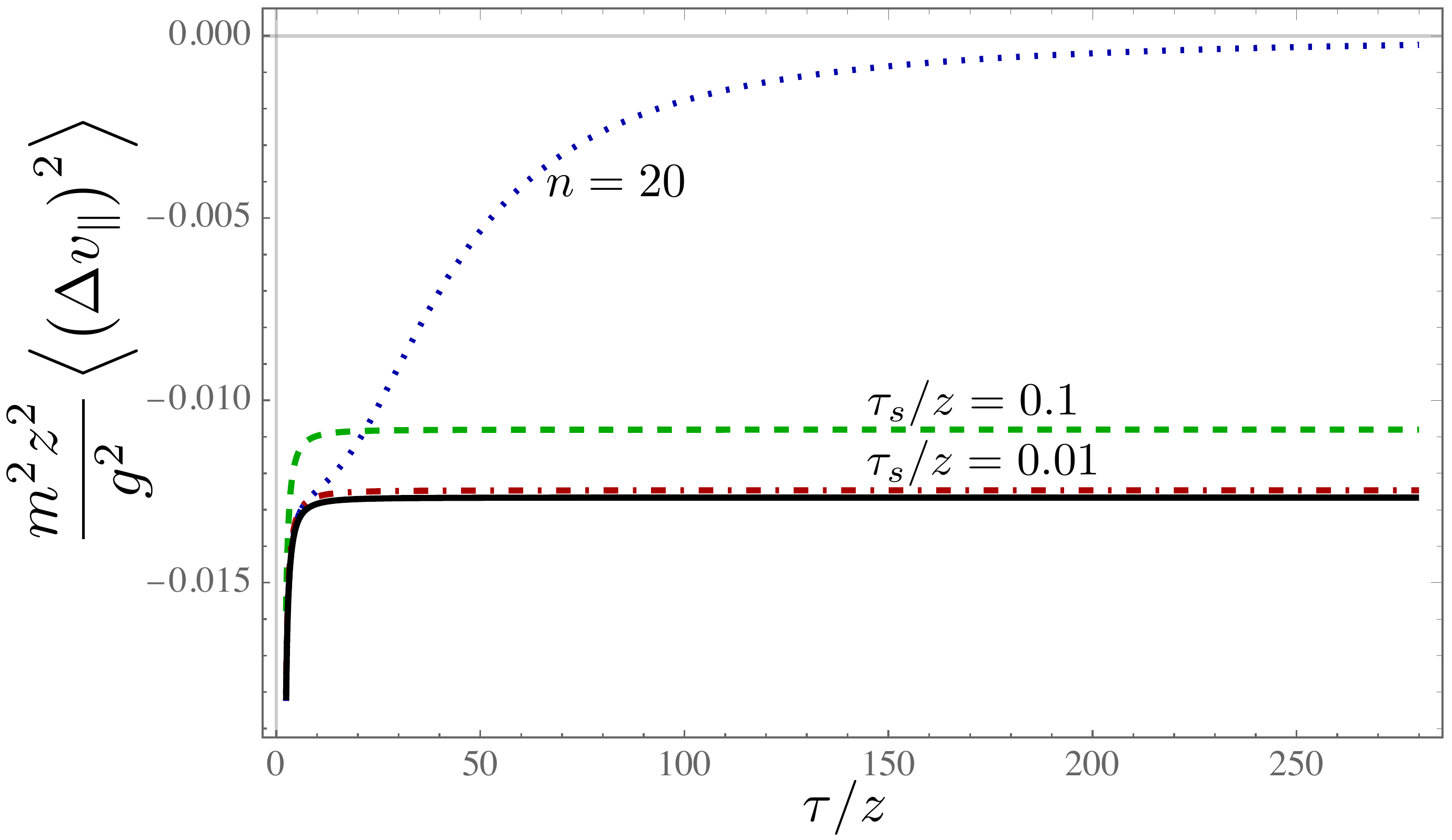}
\caption{Late time behavior of the dispersion of parallel component of the particle velocity when the switching $F_\tau^{(n)}(t)$ is implemented (dotted curve), and when the alternative switching $F_{\tau_s,\tau}(t)$ is implemented (dashed and dot-dashed curves).  The solid curve describes the sudden switching. The curves start at $\tau/z = 2.4$ in this figure.}
\label{fig5}
\end{figure}

One conclusion we get from the above discussion is that the behavior of the dispersions are closely related to both the modified vacuum state and how fast the switching between the empty space and the space presenting a reflecting wall occurs.
In order to make the above point clear let us introduce another convenient smooth switching, but now with a controllable switching time $\tau_s$. More specifically, we can select a function of $\tau$ and $\tau_s$ that keeps $\tau_s$ finite even in the limit of $\tau \rightarrow \infty$. 
In this way we  can always control the rate in which the system evolves between the two distinct states. For a finite $\tau_s$ a non null dispersion is expected to happen in the limit $\tau\rightarrow\infty$. Such scenario is implemented by using the smooth switching described by the function $F_{\tau_s,\tau}(t)=\pi^{-1}\left\{\arctan(t/\tau_s)+\arctan[(\tau-t)/\tau_s]\right\}$. This function was recently used in the study of analog quantum light cone fluctuations in a nonlinear dielectrics \cite{bessa2016}. If this switching is implemented the dispersions will present the same qualitative behaviors as those described by Eqs. (\ref{3.17}) and (\ref{3.18}), except that they will tend to a nonzero constant value at late times, which depends on the magnitude of $\tau_s$, as described in Figs. \ref{fig3} and \ref{fig5} (dashed and dot-dashed curves). Such a constant value coincides with the one given by Eq. (\ref{3.10}) when the limit of $\tau_s \rightarrow 0$ is taken, as expected. 
Detailed calculations of the quantum dispersions when this alternative switching is implemented are presented in the appendix.

\section{Kinetic energy lost by the particle due to the transition between vacuum states}
\label{e-kinetic}
We have seen that the switching time $\tau_s$ is zero in the case of (A) a sudden switching process, is a linear function of $\tau$ for a fixed $n$ when (B) the smooth switching $F_\tau^{(n)}(t)$ is implemented, and it can be arbitrarily chosen when (C) the smooth switching $F_{\tau_s,\tau}(t)$ is implemented. Furthermore, scenario A is a particular case of scenarios B or C when $n\rightarrow\infty$ or $\tau_s\rightarrow 0$ are taken, respectively. 
For a finite $\tau$, the change in the kinetic energy of the particle due to quantum vacuum fluctuations is given by the vacuum expectation value $\braket{K}=m \left( \braket{{v_x}^2}+ \braket{{v_z}^2}/2\right)$, which is mostly a negative quantity, as it can be confirmed directly by the analysis of Eqs. (\ref{3.7}) and (\ref{3.9}). Depending on how fast the switching is done, only for measuring times just after $\tau = 2 z$ the corresponding dimensionless quantity $mz^2\braket{K}/g^2$ can achieve positive values. For instance, considering the switching described by scenario B, $\braket{K}$ will always be negative for $n<12$. Larger values of $n$, which corresponds to a faster switching time, would bring a finite time interval for which this quantity becomes positive. 

\begin{figure}
\centering
\includegraphics[scale=0.4]{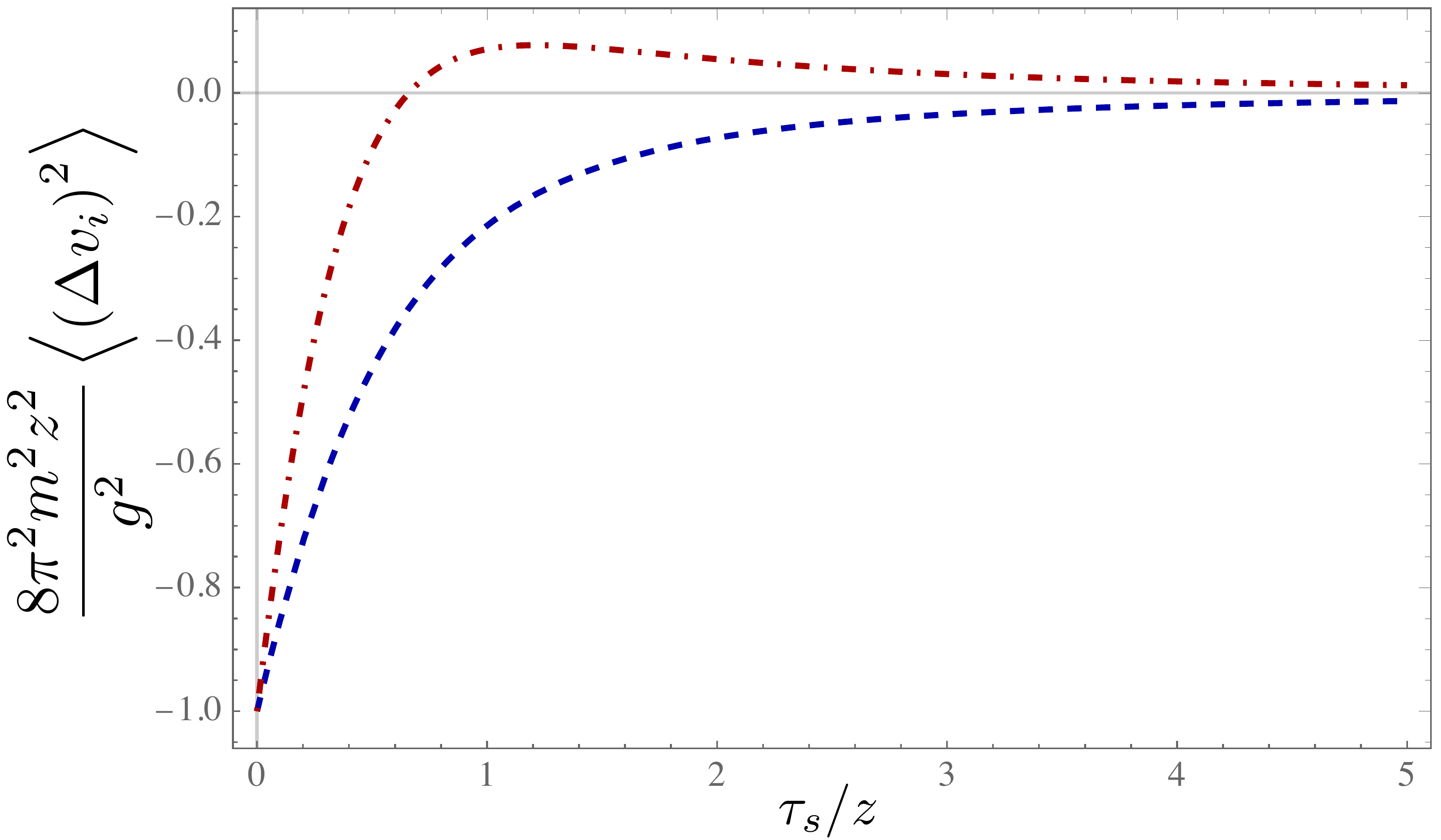}
\caption{Behavior of the dispersions of parallel (dashed curve) and perpendicular (dot-dashed curve) components of the particle velocity when the switching $F_{\tau_s,\tau}(t)$ is implemented and the regime $\tau\rightarrow\infty$ is taken. Subindex $i$ holds for parallel ($\|$) or perpendicular ($\perp$).}
\label{fig6}
\end{figure}

One important aspect that still worths analysis is the magnitude of the dispersion of the particle velocity as function of the switching time $\tau_s$ in the regime of $\tau\rightarrow\infty$. As the kinetic energy of the particle is directly related to the square of its speed, the residual effect will give us a measure of the amount of energy gained or lost by the particle due to the transition between the vacuum states of the system. In order to examine this aspect we select the switching function described in scenario C, because in this case we can directly control the rate in which the transition occurs. Hence, in the regime of $\tau/\tau_s \gg 1$,  Eqs. (\ref{3.23a}) and (\ref{3.23}) yield the following residual dispersions,
\begin{eqnarray}
\lim_{\tau\rightarrow\infty}\langle(\Delta v_\parallel)^2\rangle_{\tau_s}&=&-\frac{g^2}{8\pi^2 m^2 z^2}\left[1-\frac{\tau_s}{z}\arctan\left(\frac{z}{\tau_s}\right)\right],
\label{aa}\\
\lim_{\tau\rightarrow\infty}\langle(\Delta v_\perp)^2\rangle_{\tau_s}&=&-\frac{g^2}{8\pi^2 m^2 z^2}\left[\frac{1+2\left(\frac{\tau_s}{z}\right)^2}{1+\left(\frac{\tau_s}{z}\right)^2}-2\frac{\tau_s}{z}\arctan\left(\frac{z}{\tau_s}\right)\right].
\label{limittaus}
\end{eqnarray}
The behavior of the above residual effects are depicted in Fig. \ref{fig6}, where the dot-dashed curve is related to the perpendicular component of the particle velocity, while the dashed curve describes the behavior of its parallel component.
First, we notice that when $\tau_s \rightarrow 0$ both dispersions reduce to the sudden approximation result stated by Eq. (\ref{3.10}), as expected. This corresponds to the maximum magnitude of the residual effect. On the other hand, they go asymptotically to zero as $\tau_s/z$ increases. This is an expected result, as the predictions of both smooth switching scenarios coincide in this regime. Now, we notice that while the parallel component is a monotonically increasing function of $\tau_s/z$, the perpendicular component has a nontrivial behavior. Specifically, it cross the axis at a finite value of $\tau_s/z$. This means that if we set such specific finite switching time interval, no residual dispersion of the perpendicular component of the particle velocity will occur, in despite of the fact that in this case the transition is non adiabatic. 
At this point the effects connected to the correlations and anti correlations of the quantum scalar field cancel each other. 

\begin{figure}
\centering
\includegraphics[scale=0.4]{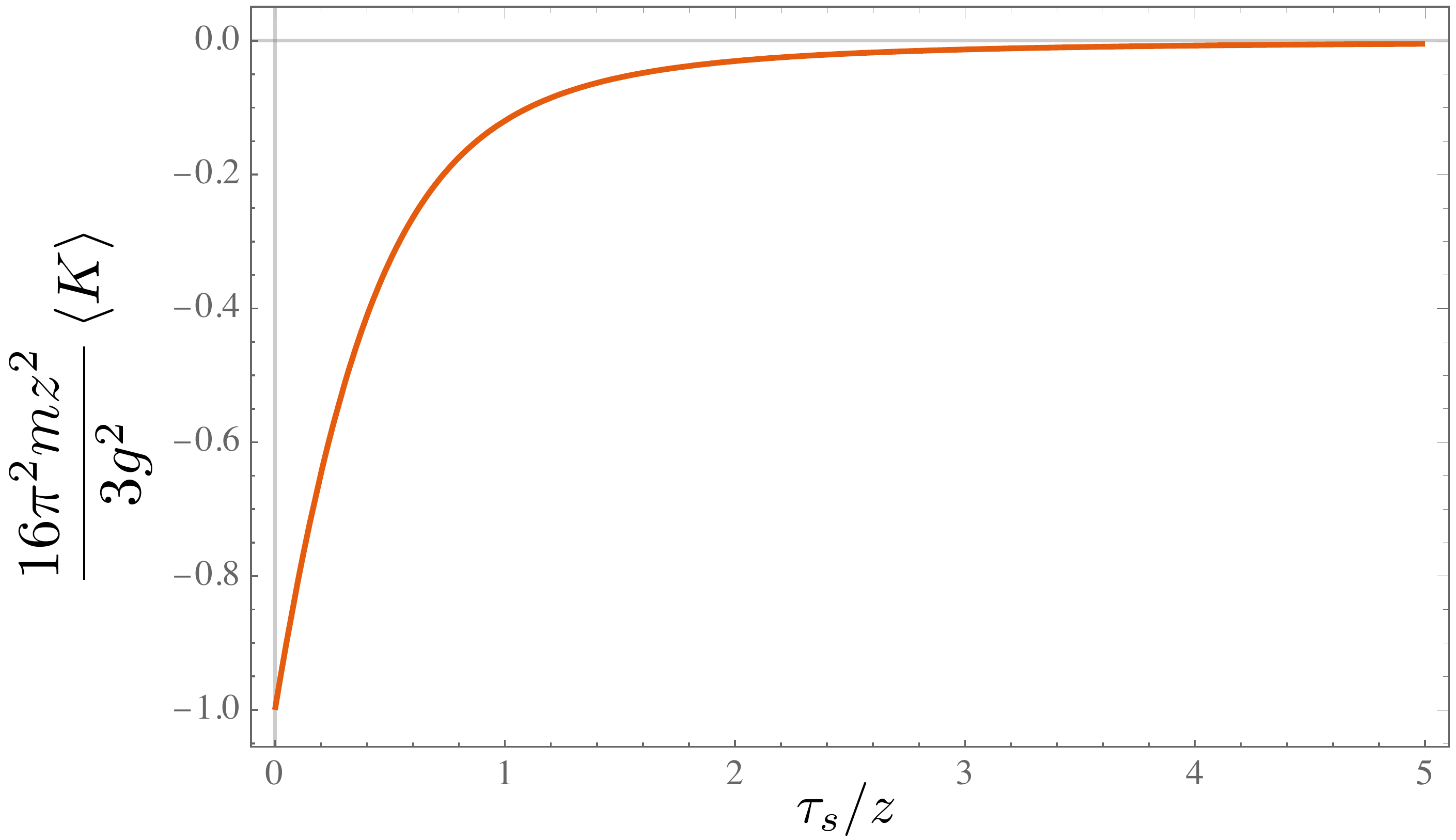}
\caption{Vacuum expectation value of the kinetic energy ($\langle K\rangle$) of the particle in the late time regime.}
\label{fig7}
\end{figure}

If only the quantum effects here studied are considered, and assuming the above late time regime,  the change in the kinetic energy of the particle is simply given by,
\begin{equation}
\braket{K}=-\frac{3 q^2}{16 \pi ^2 m z^2} \left[\frac{1+\frac{4}{3}\left(\frac{\tau_s}{z}\right)^2}{1+\left(\frac{\tau_s}{z}\right)^2}-\frac{4}{3}\frac{\tau_s}{z}\arctan\left(\frac{z}{\tau_s}\right)\right].
\label{kinetic}
\end{equation}
As depicted in Fig. \ref{fig7}, $\braket{K}$ is a increasing function of $\tau_s/z$ whose maximum magnitude occurs when the sudden process is implemented $(\tau_s \rightarrow 0)$. When $\tau_s\rightarrow \infty$ the adiabatic transition is recovered, for which no effects remain. The change in the kinetic energy of the particle is bigger for faster rate of transition, i.e., for small $\tau_s$. Notice however that in all cases $\braket{K}$ is negative, which indicates that a certain amount of kinetic energy of the particle, no matter its origin, is suppressed due to sub-vacuum fluctuations of the quantum scalar field. 

The fact that the change in the kinetic energy is non null when a transition between vacuum states occurs seems to be a clear signature of energy conservation law. However, the very mechanism behind this phenomenon can only be elucidated in a more complete model, where all classical and quantum aspects regarding the motion of the test particle and the wall should be taken into account. 

\section{Final remarks}
\label{final}
In this work the behavior of a test particle under influence of vacuum fluctuations of a real massless scalar field in the presence of a perfectly reflecting boundary was examined. Dispersions of the particle velocity were calculated in two different scenarios, corresponding to a sudden and a smooth switching processes. In the sudden scenario the dispersions exhibit localized divergences, which are linked to the over-idealized description of the system. The use of smooth switchings completely regularizes these divergences and brings some new insights on the behavior of the system, particularly in what refers to its late time regime. Gravitational and scalar field classical interactions between the particle and the wall were suppressed from our analysis, as those interactions would not contribute to the dispersions of the particle velocity. However, we should notice that thermal effects could contribute to this phenomenon. It would be interesting to compare the effects coming from thermal fluctuations with those produced by  quantum vacuum fluctuations. 

As we have seen, the change in the kinetic energy of the particle due to vacuum fluctuations is mostly a negative quantity. Moreover, in the late time regime it is always negative. A negative value of $\braket{K}$ is here understood as a manifestation of sub-vacuum fluctuations. 
We should remind that our results were obtained assuming a simplified model for which the particle position does not appreciably vary in time. Furthermore, only quantum contributions were here considered. If a more detailed model is assumed, even in the case of a particle initially at rest, the classical interaction between the particle and the reflecting wall would produce the motion of the particle perpendicularly to the wall, bringing a new (classical) term to the kinetic energy. Hence, the negative dispersion brought by quantum vacuum fluctuations would play the role of suppressing a small amount of the energy of the particle. 
When all classical and quantum effects are taken into account the total kinetic energy of the particle will surely be a positive quantity.

Our investigation was restricted to the case of a test particle interacting with a massless scalar field. Nonetheless, if a massive field is considered all results discussed in the previous sections hold as main contributions when  $Mz \ll 1$ is assumed, with $M$ denoting the field mass. Numerical analysis reveal that significant differences start to occur about $Mz \approx 1$, where the dispersions start to deviate from the massless case, but still being negative in the late time regime. When a large field mass $Mz \gg 1$ is assumed, dispersions in both directions approach zero.

Before closing, it is worth making a parallel with the similar system of a charged test particle interacting with the vacuum fluctuations of the electromagnetic field in the presence of a perfectly conducting flat wall \cite{ford2004,ford2005,hongwei2004,hongwei2006,seriu2008,seriu2009,delorenci2016}. First of all, when a sudden switching is implemented, the dispersions of the velocity components perpendicular and parallel to the wall are plagued with the same sort of divergences at $z=0$ and $\tau = 2z$ as it occurs in the scalar field case. Additionally, it was shown that the implementation of a smooth switching regularizes these divergences \cite{delorenci2016} in the same way as we have obtained for the scalar field interaction.
The behavior of the dispersions in the late time regime is also similar to the one exhibited in the scalar field case here investigated.  For instance, when the switching described by $F_\tau^{\left(n\right)}(t)$ is implemented we obtain 
$\braket{(\Delta v_\perp)^2}_{\tt electric} = -2(q/g)^2\braket{(\Delta v_\parallel)^2}_{\tt scalar}$ and $\braket{(\Delta v_\parallel)^2}_{\tt electric} \approx -2(q/g)^2\braket{\left(\Delta v_\perp\right)^2}_{\tt scalar}$, 
where $q$ denotes the electric charge of the particle. Hence, the dispersions tend to zero in the limit $\tau \rightarrow \infty$, i.e., no residual effect will appear \cite{delorenci2016}. 
Quite similar calculations using the smooth switching defined by $F_{\tau_s,\tau}(t)$ [see Eq. (\ref{int27}), in the appendix]  shows that a residual effect appears in this same limit, in agreement with our results for the scalar field. This is because the switching time $\tau_s$ can be kept finite in the late time regime. Naturally, when $\tau_s\sim\tau\rightarrow\infty$  both scenarios give the same results.
Another interesting point unveiled by the use of  $F_{\tau_s,\tau}(t)$  concerns the magnitude of the dispersions in the late time regime. In the scalar field case the residual effect corresponding to the sudden process sets a lower limit to the dispersions, as shown in Fig. \ref{fig3}.  On the other hand, when the electromagnetic case is considered we will have an opposite behavior, and the sudden process sets an upper limit to the dispersions. In both cases the sudden process is achieved in the limit of $\tau_s \rightarrow 0$.
Finally, we notice that while the kinetic energy of the particle is always lessen in the case of the scalar field, in the electromagnetic case an amount of kinetic energy is possible to be gained by the charged particle due to the transition process between the vacuum states. 
However, depending on how fast the transition occurs, the change in the kinetic energy of the particle can be negative even in the electromagnetic case.
This is an issue that deserves a detailed investigation.

\appendix
\section{Detailed calculations using $F_{\tau_s,\tau}\left(t\right)$}
\begin{figure}
\centering
\includegraphics[scale=0.325]{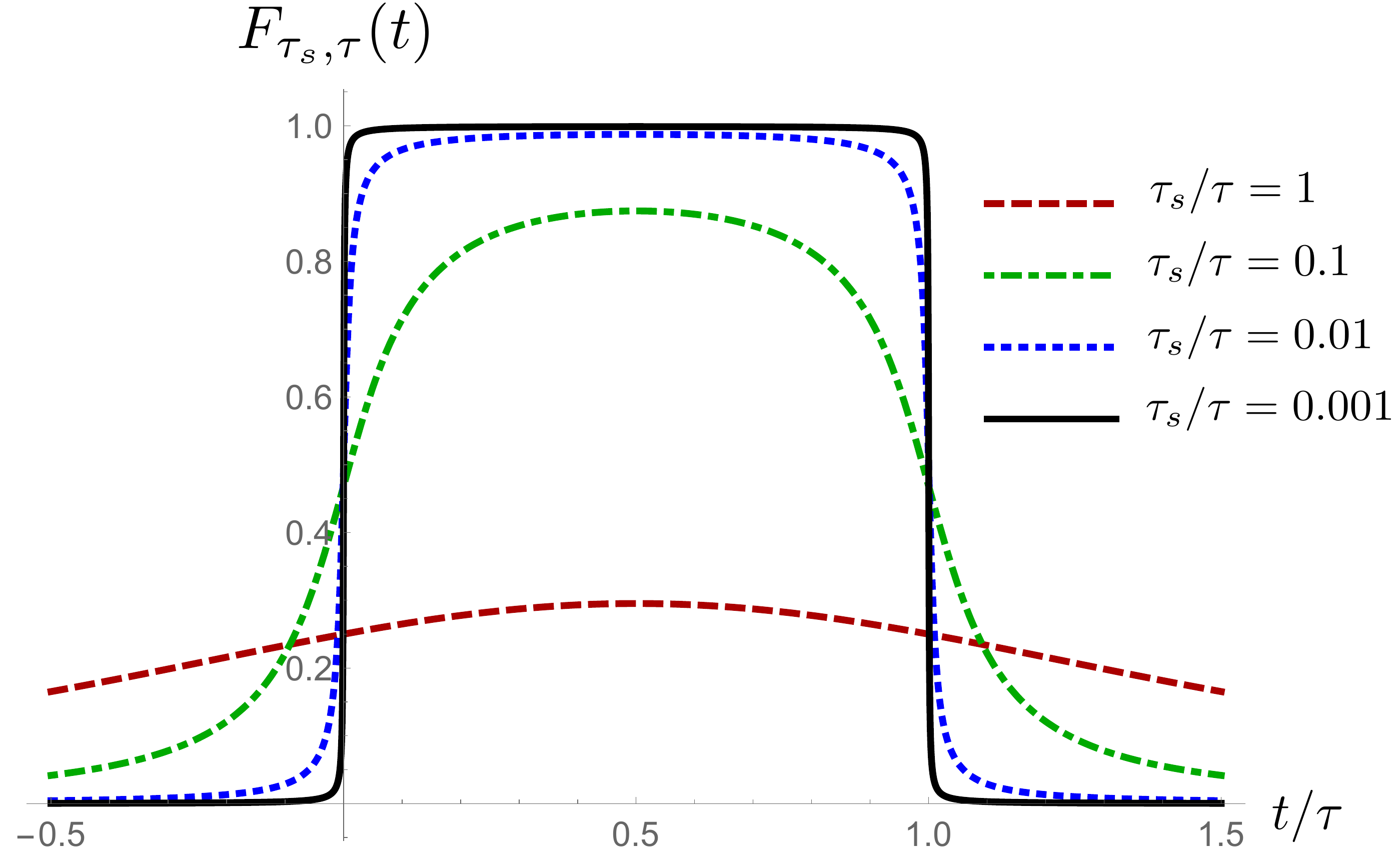}
\caption{Behavior of $F_{\tau_s,\tau}\left(t\right)$ given by Eq.~\eqref{int27} for some representative value of $\tau_s/\tau$. Here the sudden switching regime is achieved when $\tau_s/\tau \rightarrow 0$. The switching function described by $F_\tau^{\left(n\right)}(t)$ presents a quite similar behavior, but with the sudden switching being approached in the regime of $n>>1$.}
\label{ftts}
\end{figure}

Let the smooth switching be defined by  
\begin{equation}
F_{\tau_s,\tau}\left(t\right)=\frac{1}{\pi}\left[\arctan{\left(\frac{t}{\tau_s}\right)}+\arctan{\left(\frac{\tau-t}{\tau_s}\right)}\right],
\label{int27}
\end{equation}
in which the switching time $\tau_s$ appears explicitly as a parameter. Its behavior for several values of the rate $\tau_s/\tau$ is depicted in Fig. \ref{ftts}, which is essentially the same behavior exhibited by the switching function $F^{(n)}_{\tau}\left(t\right)$. The sudden switching is reproduced in the limit $\tau_s\rightarrow 0$, for a fixed $\tau$.

The dispersions of the particle velocity are obtained following the same steps we used in section \ref{sss}. Hence, introducing the switching function given by Eq.~\eqref{int27} in Eq.~\eqref{int10a},  using the integral representation of $G_{{}_R}^{{}^{(1)}}({\bf x},t;{\bf x}',t')$,
\begin{equation}
G_{{}_R}^{{}^{(1)}}({\bf x},t;{\bf x}',t')=-\frac{1}{2\pi^2\sigma}\int_0^\infty e^{-ik\left(t-t^\prime\right)}\sin{\left(k\sigma\right)}dk,
\label{3.19}
\end{equation}
with $\sigma=\sqrt{(x-x')^2+(y-y')^2+(z+z')^2}$,  and also the formula
\begin{equation}
\int_{-\infty}^\infty\int_{-\infty}^\infty dt dt'F_{\tau_s,\tau}(t)F_{\tau_s,\tau}(t')e^{-ik (t-t')}=\frac{2[1-\cos(k\tau)]e^{-2k \tau_s}}{k^2},
\end{equation}
we obtain, 
\begin{equation}
\braket{\left(\Delta v_i\right)^2}_{\tau_s}=-\frac{g^2}{2\pi^2m^2}\left[\frac{\partial}{\partial x_i}\frac{\partial}{\partial x_i^\prime}\frac{1}{\sigma}\int_0^\infty d k\sin{\left(k\sigma\right)}\frac{e^{-2k\tau_s}}{k^2}\left[1-\cos{\left(k\tau\right)}\right]\right]_{\textbf{x}^\prime=\textbf{x}}.
\label{3.21}
\end{equation}

If we use the exponential representation of the trigonometric functions, the above integral is solved by means of \cite{gradshteyn} (eq. 8.350.2)
$$\int_0^\infty\frac{e^{-\lambda k}}{k^2}dk=\lambda\lim_{\epsilon\to0}\int_{\lambda\epsilon}^\infty\frac{e^{-\alpha}}{\alpha^2}d\alpha=\lambda\lim_{\epsilon\to0}\Gamma\left(-1,\lambda\epsilon\right),$$
where, $\Gamma\left(-1,\lambda\epsilon\right)$ is the incomplete Gamma function, whose expansion for small values of $\epsilon$ reads $\Gamma\left(-1,\lambda\epsilon\right)\approx (\lambda\epsilon)^{-1} -1+\gamma+\ln{\lambda\epsilon},$ with $\gamma$ the Euler-Mascheroni constant.

Finally, after solving the integral, operating with the derivatives, and taking the limit of point coincidence, we obtain the following closed expressions for the dispersions,
\begin{eqnarray}
%\begin{aligned}
\langle(\Delta v_\parallel)^2\rangle_{\tau_s}&=&
\frac{g^2}{64\pi^2m^2z^3}\left\{\tau\ln\left[\frac{(\tau-2z)^2+4\tau_s^2}{(\tau+2z)^2+4\tau_s^2}\right]+4\tau_s\arg\left[1+\frac{\tau^2}{4(\tau_s-iz)^2}\right]\right\},
\label{3.23a}\\
\langle(\Delta v_\perp)^2\rangle_{\tau_s}&=&2\langle(\Delta v_\parallel)^2\rangle_{\tau_s}+\frac{g^2\tau^2}{8\pi^2m^2}\frac{\tau^2-4z^2+12\tau_s^2}{(\tau_s^2+z^2)[(\tau-2z)^2+4\tau_s^2][(\tau+2z)^2+4\tau_s^2]},
%\end{aligned}
\label{3.23}
\end{eqnarray}
where the function $\arg\left(r\right)$ is the argument of the complex number $r$.

\acknowledgments
This work was partially supported by the Brazilian research agencies CNPq (Conselho Nacional de Desenvolvimento Cient\'{\i}fico e Tecnol\'ogico) under grant 302248/2015-3, FAPEMIG (Funda\c{c}\~ao de Amparo \`a Pesquisa do Estado de Minas Gerais),  FAPESP (Funda\c{c}\~ao de Amparo \`a Pesquisa do Estado de S\~ao Paulo) under grant 2015/26438-8, and CAPES (Coordena\c{c}\~ao de Aperfei\c{c}oamento de Pessoal de N\'{\i}vel Superior).


\begin{thebibliography}{99}

\bibitem{lamoreaux2005}
S. K. Lamoreaux,
{\it The Casimir force: background, experiments, and applications,}
{\it Rep. Prog. Phys.} {\bf 68} (2005) 201.

\bibitem{klimchitskaya2009}
G. L. Klimchitskaya, U. Mohideen, and V. M. Mostepanenko,
{\it The Casimir force between real materials: experiment and theory,}
{\it Rev. Mod. Phys.} {\bf 81} (2009) 1827.

\bibitem{liu2016}
X. Liu, Y. Li, and H. Jing,
{\it Casimir switch: steering optical transparency with vacuum forces,}
{\it Sci. Rep.}  {\bf 6} (2016) 27102.

\bibitem{delorenci2015xi}
V. A. De Lorenci, L. G. Gomes and E. S. Moreira Jr., 
{\it Hot scalar radiation setting bounds on the curvature coupling parameter,}
{\it Class. Quant. Grav.} {\bf 32} (2015) 085002.
	
\bibitem{moreira2017}
E. S. Moreira, 
{\it Hot scalar radiation around a cosmic string setting bounds on the coupling parameter $\xi$,}
{\it JHEP} {\bf 03} (2017) 105. 

\bibitem{delorenci2004}
V. A. De Lorenci, R. Klippert, and E. S. Moreira, Jr.,
{\it Semiclassical backreaction around a cosmic dislocation,}
{\it Phys. Rev. D} {\bf 71} (2005) 024005.

\bibitem{ford2004} 
H. Yu and L. H. Ford, 
{\it Vacuum fluctuations and Brownian motion of a charged test particle near a reflecting boundary,}
{\it Phys. Rev. D}, {\bf 70} (2004) 065009.

\bibitem{ford2005}
L. H. Ford, 
{\it Stochastic Spacetime and Brownian Motion of Test Particles,} 
{\it Int. J. Theor. Phys.} {\bf 44} (2005) 1753.

\bibitem{hongwei2004} 
H. Yu and J. Chen, 
{\it Brownian motion of a charged test particle in vacuum between two conducting plates,}
{\it Phys. Rev. D} {\bf 70} (2004) 125006.

\bibitem{hongwei2006} 
H. Yu, J. Chen, and P. Wu, 
{\it Brownian motion of a charged test particle near a reflecting boundary at finite temperature,}
{\it JHEP} {\bf 02} (2006) 058.

\bibitem{parkinson2011}
V. Parkinson and L. H. Ford, 
{\it Model for noncancellation of quantum electric field fluctuations,} 
{\it Phys. Rev. A} {\bf 84} (2011) 062102.

\bibitem{seriu2008} 
M. Seriu and C. H. Wu, 
{\it Switching effect on the quantum Brownian motion near a reflecting boundary,}
{\it Phys. Rev. A} \textbf{77} (2008) 022107.

\bibitem{seriu2009} 
M. Seriu and C. H. Wu, 
{\it Smearing effect due to the spread of a probe particle on the Brownian motion near a perfectly reflecting boundary,}
{\it Phys. Rev. A} \textbf{80} (2009) 052101.

\bibitem{delorenci2016}
V.  A. De Lorenci, C. C. H. Ribeiro, and M. M. Silva,
{\it Probing quantum vacuum fluctuations over a charged particle near a reflecting wall,}
{\it Phys. Rev. D} {\bf 94} (2016) 105017.

\bibitem{bessa2009}
C. H. G. Bessa, V. B. Bezerra, and L. H. Ford,
{\it Brownian motion in Robertson-Walker spacetimes from electromagnetic vacuum fluctuations,}
{\it J. Math. Phys.} {\bf 50} (2009) 062501.

\bibitem{delorenci2014}
V.  A. De Lorenci, E. S. Moreira, Jr., and M. M. Silva, 
{\it Quantum Brownian motion near a point-like reflecting boundary,}
{\it Phys. Rev. D} {\bf 90} (2014) 027702.

\bibitem{bartolo2015}
N. Bartolo, S. Butera, M. Lattuca, R. Passante, L. Rizzuto, and S. Spagnolo,
{\it Vacuum Casimir energy densities and field divergences at boundaries,}
{\it J. Phys.: Condens. Matter} {\bf 27} (2015) 214015.

\bibitem{wang2014}
Q. Wang and W. G. Unruh, 
{\it Motion of a mirror under infinitely fluctuating  quantum vacuum stress,}
{\it Phys. Rev. D} {\bf 89} (2014) 085009.

\bibitem{bessa2017}
C. H. G. Bessa, V. B. Bezerra, E. R. Bezerra de Melo, and H. F. Mota,
{\it Quantum Brownian motion in an analog Friedmann-Robertson-Walker geometry,}
{\it Phys. Rev. D} {\bf 95} (2017) 085020.

\bibitem{birrel1982} 
N. D. Birrel and P. C. W. Davies, {\em Quantum Fields in Curved Space},
Cambridge University Press, Cambridge (1982), Secs. 2.1 and 4.3.

\bibitem{johnson2002}
P. R. Johnson and B. L. Hu, 
{\it Stochastic theory of relativistic particles moving in a quantum field: scalar Abraham-Lorentz-Dirac-Langevin equation, radiation reaction, and vacuum fluctuations,} 
{\it Phys. Rev. D} {\bf 65} (2002) 065015.

\bibitem{ford1998}
L. H. Ford and N. F. Svaiter, 
{\it Vacuum energy density near fluctuating boundaries,}
{\it Phys. Rev. D} {\bf 58}, 065007 (1998).

\bibitem{armata2015}
F. Armata and R. Passante, 
{\it Vacuum energy densities of a field in a cavity with a mobile boundary,}
{\it Phys. Rev. D} {\bf 91}, 025012 (2015).

\bibitem{bessa2016}
C. H. G. Bessa, V. A. De Lorenci, L. H. Ford, and C. C. H. Ribeiro, 
{\it Model for lightcone fluctuations due to stress tensor fluctuations,}
{\it Phys. Rev. D} {\bf 93} (2016) 064067.

\bibitem{gradshteyn}
I. S. Gradshteyn and I. M. Ryzhik,
{\it Table of Integrals, Series, and Products}, Academic Press, New York, (2007). %, Eq. 8.350.2. (1.411.3; 1.511.1; 1.622.3; 1.643.1)

\end{thebibliography}
\end{document}